\documentclass[twocolumn,secnumarabic,nobibnotes,reprint,superscriptaddress,amssymb,amsmath,aps,showpacs,10pt,floatfix,prb,longbibliography]{revtex4-2}

\usepackage{ulem}
\usepackage{graphicx}
\usepackage{amsmath}
\usepackage{amsfonts}
\usepackage{stmaryrd}
\usepackage{bbm,bm}
\usepackage{mathrsfs}
\usepackage{tipa}
\usepackage{amssymb}
\usepackage{txfonts}
\usepackage{dcolumn}
\usepackage{epstopdf}
\usepackage{float}
\usepackage{array}
\usepackage{longtable}
\usepackage{subfigure}
\makeatletter
\renewcommand{\thesubfigure}{\alph{subfigure}}
\renewcommand{\@thesubfigure}{\thesubfigure}
\renewcommand{\p@subfigure}{\thefigure}
\makeatother
\usepackage{url}
\usepackage[colorlinks,bookmarks=true,citecolor=blue,linkcolor=blue,urlcolor=blue]{hyperref} 

\begin{document}
	
	\title{Efficient photo-Nernst terahertz emission in single heavy-metal films}
	
	\author{Lei Wang}
	\altaffiliation{These authors contributed equally to this work.}
	\affiliation{Beijing National Laboratory for Condensed Matter Physics, Institute of Physics, Chinese Academy of Sciences, Beijing 100190, China}
	\affiliation{Center for Advanced Quantum Studies and Department of Physics, Beijing Normal University, Beijing 100875, China}
	
	\author{Linxuan Song}
	\altaffiliation{These authors contributed equally to this work.}
	\affiliation{Beijing National Laboratory for Condensed Matter Physics, Institute of Physics, Chinese Academy of Sciences, Beijing 100190, China}
	\affiliation{Institute of Quantum Materials and Devices, School of Electronics and Information Engineering, Tiangong University, Tianjin 300387, China}
	
	\author{Elbert E. M. Chia}
	\affiliation{Division of Physics and Applied Physics, School of Physical and Mathematical Sciences, Nanyang Technological University, Singapore 637371, Singapore}
	
	\author{Peijie Sun}
	\affiliation{Beijing National Laboratory for Condensed Matter Physics, Institute of Physics, Chinese Academy of Sciences, Beijing 100190, China}
	\affiliation{School of Physical Sciences, University of Chinese Academy of Sciences, Beijing 100190, China}
	
	\author{Jianlin Luo}
	\affiliation{Beijing National Laboratory for Condensed Matter Physics, Institute of Physics, Chinese Academy of Sciences, Beijing 100190, China}
	\affiliation{School of Physical Sciences, University of Chinese Academy of Sciences, Beijing 100190, China}
	
	\author{Rongyan Chen}
	\email{rychen@bnu.edu.cn}
	\affiliation{Center for Advanced Quantum Studies and Department of Physics, Beijing Normal University, Beijing 100875, China}
	
	\author{Yong-Chang Lau}
	\email{yongchang.lau@iphy.ac.cn}
	\affiliation{Beijing National Laboratory for Condensed Matter Physics, Institute of Physics, Chinese Academy of Sciences, Beijing 100190, China}
	\affiliation{School of Physical Sciences, University of Chinese Academy of Sciences, Beijing 100190, China}
	
	\author{Xinbo Wang}
	\email{xinbowang@iphy.ac.cn}
	\affiliation{Beijing National Laboratory for Condensed Matter Physics, Institute of Physics, Chinese Academy of Sciences, Beijing 100190, China}
	\affiliation{School of Physical Sciences, University of Chinese Academy of Sciences, Beijing 100190, China}

	\date{\today}

	\begin{abstract} 
		State-of-the-art metallic terahertz (THz) emitters rely predominantly on spintronic heterostructures, where heavy metals serve as passive spin-to-charge converters. Here, we demonstrate efficient THz radiation from standalone Pt nanofilms at cryogenic temperatures and under external magnetic fields. The governing mechanism is identified as the ultrafast photo-Nernst effect, wherein a transient thermal gradient drives a transverse charge current. The THz emission polarity is directly dictated by the sign of the Nernst coefficient, as verified by the phase reversal observed between Pt and W or Ta. Remarkably, both thickness scaling and alloying-induced suppression of thermal conductivity independently amplify the single-layer emission to levels comparable with benchmark spintronic bilayers. These findings redefine the established role of heavy metals from passive spin-sinks to active THz emitters, uncovering a universal emission paradigm applicable across diverse spintronic and quantum materials.
	\end{abstract}
	\maketitle
	
	\setlength{\parindent}{2em}
	\section{Introduction}
	Spintronic terahertz emitters based on ferromagnetic/non-magnetic (FM/NM) heterostructures have emerged as efficient broadband THz sources, attracting increasing attention in both fundamental condensed-matter research and ultrafast THz photonics\cite{kampfrath2013terahertz, Seifert2016Efficient, Cheng21Studying, Seifert22Spintronic, Kumar23Ultrafast}. Under femtosecond optical excitation, an ultrafast spin current is launched in the FM layer and injected into the adjacent NM layer, where the inverse spin Hall effect (ISHE) converts it into a transverse charge current that emits THz radiation. The versatility of this spintronic framework has been demonstrated through various materials, where the spin-generating layers span ferromagnets\cite{kampfrath2013terahertz,Seifert2016Efficient,Wu17High} to antiferromagnets\cite{Qiu21Ultrafast,Huang22AFM, rongione2023emission,hamara2024FeRh, Metzger25Separating, ChengYu2025antiferromagnet} and ferrimagnets\cite{Huisman17Spin, Fix20Thermomagnetic}, and the spin-conversion layers include heavy metals,\cite{kampfrath2013terahertz,Seifert2016Efficient,Wu17High} topological materials,\cite{Wang18Ultrafast,Rongione22SBT,Tong21Enhanced,Chen21Efficient} Rashba interfaces\cite{Zhou18Broadband,Control18Jungfleisch} and semiconductors\cite{Cheng19MoS2,Comstock23GaN}. Within this paradigm, heavy metals such as Pt have been frequently employed  as an efficient but passive spin-to-charge converters. Consequently, a standalone Pt film, without an adjacent spin-source layer, has been considered incapable of generating THz radiation.
	
	Distinct from interfacial spin-conversion geometries, alternative pathways for THz generation highlight the importance of ultrafast thermally driven processes. In single-layer ferromagnetic films, the anomalous Nernst effect (ANE), where a transient temperature gradient induced by hot-electron thermalization drives a transverse charge current, has recently been established as a dominant emission mechanism~\cite{Zhang23FM, Feng23NL}. For non-magnetic materials, generating transverse thermoelectric currents requires breaking spatial symmetry. This is achieved either through intrinsic structural anisotropy, which sustains spontaneous zero-field photothermoelectric currents \cite{Yordanov23Generation}, or via asymmetric sample geometries. For instance, a thickness gradient creates an in-plane thermal asymmetry that couples with a weak external magnetic field to drive the ordinary Nernst effect, as demonstrated for THz generation in Dirac semimetals \cite{LuWei2022ultrafast}. 
	
	However, these prior observations of transverse thermoelectric THz radiation are confined to quantum materials hosting giant intrinsic transverse transport coefficients. In conventional non-magnetic metals, the corresponding steady-state ordinary Nernst response is severely suppressed by the Sondheimer cancellation within the Mott regime, rendering the transverse charge current negligible \cite{Behnia2016Nernst,Luo23Nernst,Chiang24Nernst}. Recently, high external magnetic fields have unveiled previously inaccessible ultrafast magneto-transport regimes, as evidenced by the field-induced ultrafast demagnetization of canted antiferromagnetic sublattices in Pt/NiO heterostructures \cite{Metzger25Separating} and the dynamic magneto-chiral responses in elemental Te \cite{Huang2026Te}. Consequently, a critical question emerges: can the coupling of a sub-picosecond thermal gradient with a high magnetic field bypass this steady-state thermodynamic limit to drive an ultrafast Nernst current in prototypical heavy metals?

	\begin{figure*}
		\includegraphics[width=0.75\linewidth]{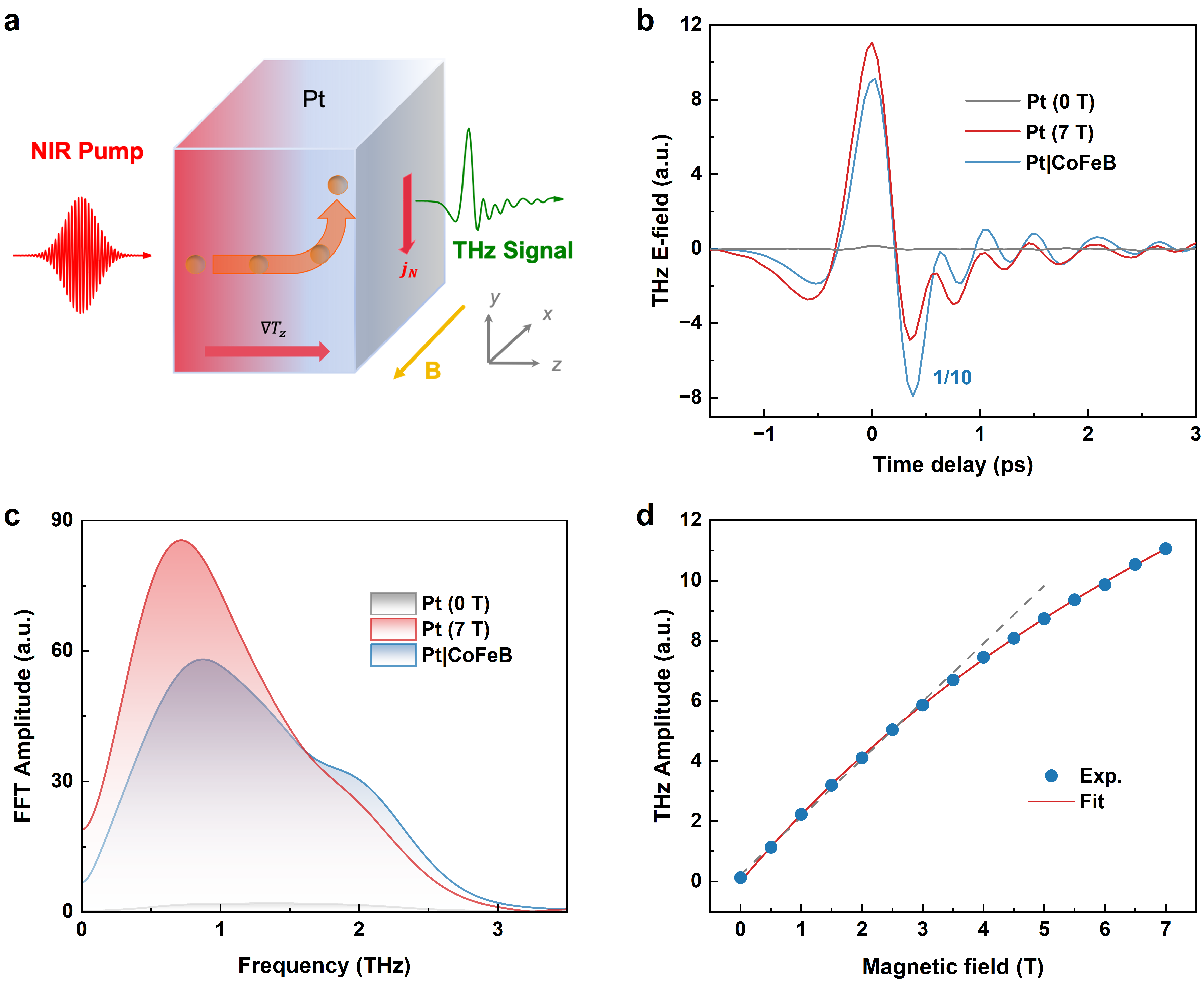}
		\centering		\vspace{-0.2cm}
		\subfigure{\label{1a}} 
		\subfigure{\label{1b}}
		\subfigure{\label{1c}}
		\subfigure{\label{1d}} 
		\caption{\textbf{Magnetic-field-induced THz emission from a single Pt film.} \textbf{a} Schematic of the experimental geometry for THz emission spectroscopy. \textbf{b} Representative time-domain THz waveforms emitted from the single Pt film at 10 K under external magnetic fields, alongside a reference signal from a Pt(3 nm)/CoFeB(3 nm) heterostructure scaled by a factor of 0.1 for amplitude comparison. \textbf{c} Corresponding Fourier-transformed amplitude spectra of the THz transients in \textbf{(b)}. \textbf{d} THz peak amplitude as a function of external magnetic field at 10 K. The solid line indicates the fit based on the semiclassical Drude-Boltzmann transport model. The dashed line represents a linear fit to the low-field data ($<$ 3 T), extrapolated to highlight the sublinear deviation.} \label{1}
	\end{figure*}
		
	To address this question, efficient THz radiation is directly observed from standalone Pt nanofilms at 10 K and 7 T. By systematically investigating the emission symmetries, material dependence, and temperature evolution, the ultrafast photo-Nernst effect (PNE) is identified as the physical origin. The thermo-magnetic mechanism is further validated by a direct correlation between the emitted THz polarity and the inherent Nernst coefficients across different heavy metals, specially Pt, W, and Ta. Furthermore, the emission efficiency is substantially amplified by suppressing the lattice thermal conductivity via Pt-Ti alloying. For bare Pt, the 2.6-nm optimal thickness arises from a direct trade-off between pump absorption and THz conductive screening. Our results extend beyond the conventional paradigm of heavy metals acting solely as passive spin-sinks, establishing their intrinsic capability for active THz generation driven by transient magneto-thermoelectric dynamics.
	
	\section{Results and discussion}
	\noindent\textbf{Magnetic-field-induced THz emission in standalone Pt.} The THz emission measurements are performed in a standard transmission geometry (Fig. \ref{1a}). A 5-nm-thick single Pt nanofilm, deposited onto a MgO substrate (Supplementary Fig. S2), is excited by femtosecond near-infrared pulses at normal incidence along the $z$-axis. With an in-plane magnetic field applied along the $x$-axis, the vertical THz electric field ($E_y$) is detected via  electro-optic sampling (Supplementary Fig. S1). As shown in Fig. \ref{1b}, a pronounced single-cycle THz pulse is generated from the standalone Pt film at 10 K and 7 T. At zero magnetic field, the signal collapses to a negligible background originating from the parasitic emission of the cryostat windows (Supplementary Fig. S3). Under identical experimental conditions, the peak-to-peak THz amplitude of the single-layer Pt reaches approximately 10\% of the emission strength of an thickness-optimized Pt(3 nm)/CoFeB(3 nm) spintronic emitter~\cite{Seifert2016Efficient}. The corresponding fast Fourier transform amplitude spectrum (Fig. \ref{1c}) reveals a bandwidth extending up to 3 THz. This spectral range is comparable to that of reference heterostructures, primarily constrained by the bandwidth of the ZnTe crystal.
	
	\begin{figure*}
		\includegraphics[width=0.75\linewidth]{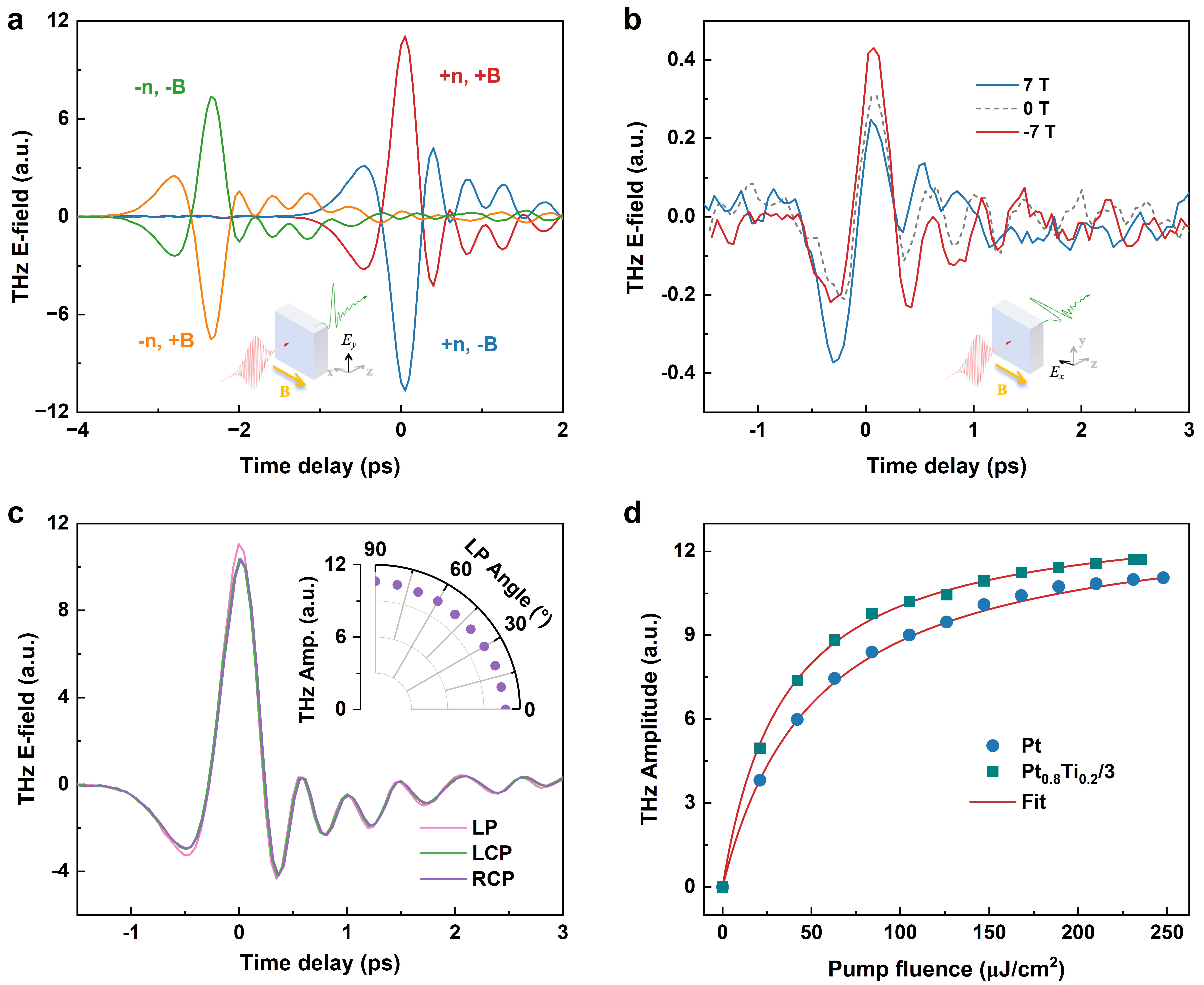}
		\centering \vspace{-0.2cm}
		\subfigure{\label{2a}} 
		\subfigure{\label{2b}}
		\subfigure{\label{2c}}
		\subfigure{\label{2d}}
		\caption{\textbf{Symmetry signatures and pump-fluence dependence of the THz emission.} \textbf{a} Time-domain THz waveforms measured at 10 K for opposite magnetic-field directions, with the optical pump incident from the film side (+n) and substrate side (-n). The THz polarity reverses upon flipping either the magnetic field or the excitation geometry. \textbf{b} The THz electric field ($E_x$) detected parallel to $\bm{B}$ at 10 K and 7 T, revealing a negligible THz emission. \textbf{c} THz transients generated by linearly polarized (LP), right-circularly polarized (RCP), and left-circularly polarized (LCP) pump pulses. The inset displays the THz amplitude as a function of the linear polarization angle from $0^\circ$ to $90^\circ$. \textbf{d} Pump fluence dependence of THz emission for bare Pt film and the Pt$_{0.8}$Ti$_{0.2}$ alloy. The THz amplitude of the alloy is scaled down by a factor of 3 for clarity. The solid lines indicate fits to the empirical saturation model.} \label{2}
	\end{figure*}
		
	To elucidate the underlying emission mechanism, we investigate the magnetic field dependence of the THz emission. As depicted in Fig. \ref{1d}, the THz peak amplitude increases linearly at low fields but develops a slight sublinear deviation above 3 T (Supplementary Fig. S4a). The field-dependent behavior is well captured by the semiclassical Drude-Boltzmann transport model\cite{Behnia2016Nernst,LuWei2022ultrafast}, where the radiated field follows $E_{\mathrm{THz}}(B)\propto B/[1+\left(\mu_{\mathrm{eff}} B\right)^{2}]$. By fitting the experimental data, an effective carrier mobility $\mu_{\mathrm{eff}}$ of 0.087 $\mathrm{m^{2}/Vs}$ is extracted. This transient mobility surpasses the steady-state DC value of sputtered Pt films by several orders of magnitude~\cite{Fischer1980}, indicating that the ultrafast transport of these hot carriers drives the THz generation prior to their complete thermalization with the lattice \cite{Cheng19MoS2}.

	\vspace{0.2cm} \noindent\textbf{Symmetry signatures of the THz emission.}	A symmetry analysis under varying excitation geometries (Fig. \ref{2}) was conducted to disentangle the exact physical origin from competing mechanisms. As shown in Fig. \ref{2a},  a polarity reversal of the transverse THz component ($E_y$) is observed upon either reversing the in-plane magnetic field or flipping the sample to alternate the pump-incidence direction (Pt-side versus substrate-side illumination). The temporal shift between the waveforms arises from the difference in the refractive indices of the MgO substrate at near-infrared and THz frequencies. In contrast, the horizontal THz component parallel to the magnetic field ($E_x$, Fig. \ref{2b}) remains negligible and insensitive to field reversal. These observations confirm that the emitted THz radiation is linearly polarized, with its transient electric field orthogonal to the applied magnetic field.
	
	\begin{figure*}
		\includegraphics[width=0.8\linewidth]{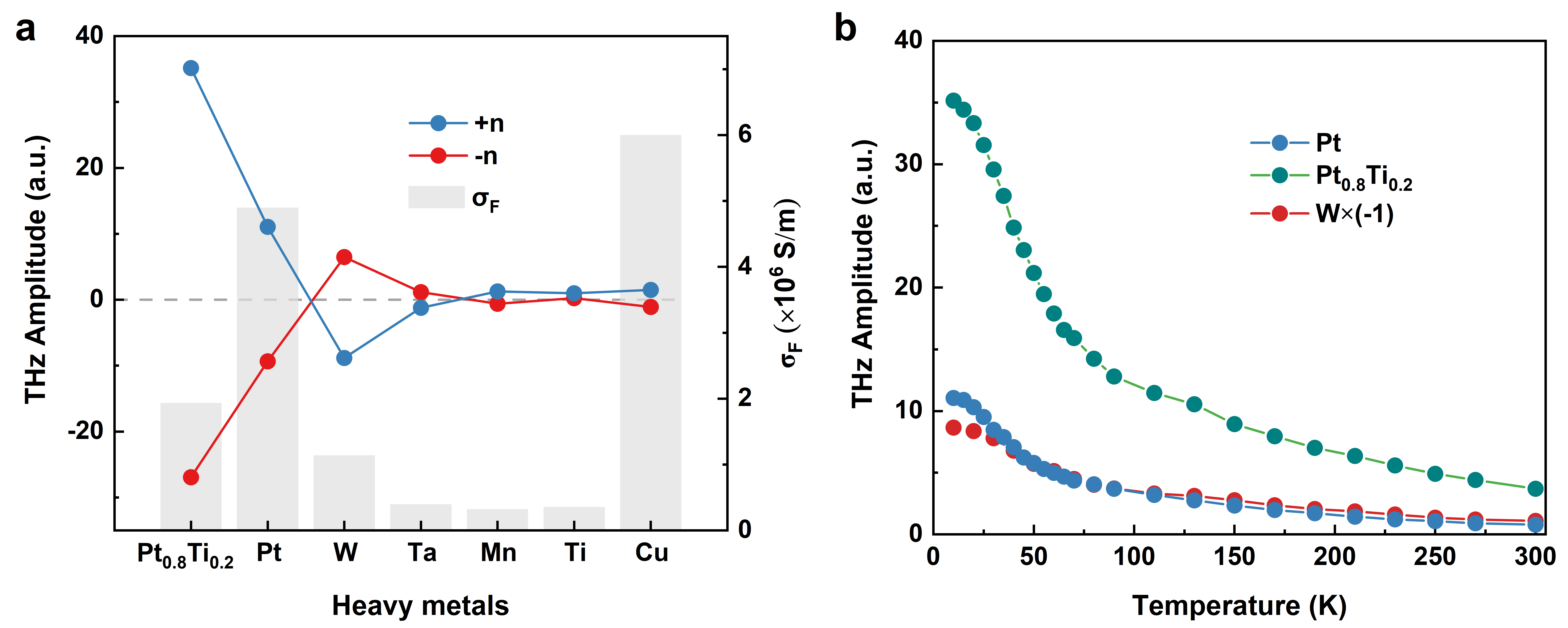}  \centering \vspace{-0.2cm}
		\subfigure{\label{3a}} 
		\subfigure{\label{3b}}
		\caption{\textbf{Material dependence and thermal evolution of the photo-Nernst emission.} \textbf{a} THz amplitudes from various 5-nm-thick single heavy-metal films measured at 10 K and 7 T. Red and blue symbols correspond to front-side and back-side optical pump excitation, respectively. The gray bars represent the electrical conductivity of each metal. \textbf{b} Temperature dependence of the THz amplitude for  5-nm-thick Pt, Pt$_{0.8}$Ti$_{0.2}$ alloy and W films. The amplitude for W is scaled by a factor of -1 for direct comparison.}  \label{3}
	\end{figure*}
	
	Another defining signature is the pump-polarization independence of the THz emission. As shown in Fig. \ref{2c}, the THz peak amplitude remains invariant upon continuous rotation of the linear polarization axis, and identical transients are detected under right- and left-circularly polarized pump excitations. The polarization insensitivity precludes helicity-driven mechanisms, such as the circular photogalvanic effect, alongside other parasitic nonlinear optical processes \cite{pettine2023ultrafast,Emerging2024wu2D}. Furthermore, the THz amplitude increases linearly with the pump fluence, but a pronounced saturation dominates at higher excitation densities (Fig. \ref{2d} and Supplementary Fig. S4b). The measured pump-fluence dependence is reproduced by the empirical model $E_{\mathrm{THz}}$ $\propto$ $F / (F + F_{\mathrm{sat}})$, from which a saturation fluence of $F_{\mathrm{sat}}$ $\simeq$ 52.5 $\mu$J/cm$^2$ is extracted. Such a low value contrasts with the mJ/cm$^2$-scale saturation observed in benchmark spintronic bilayers\cite{Zhang2018JPD,Jin2023AdvPhysRes}, signifying a distinct driving mechanism within the standalone Pt film.
	
		These emission features, specifically the $\bm{E} \perp \bm{B}$ radiation geometry, the polarity reversals, and the pump-polarization insensitivity, strictly align with the symmetry requirements of the ISHE in conventional spintronic emitters.~\cite{Cheng21Studying, Seifert22Spintronic, Kumar23Ultrafast}. Whereas this mechanism mediates the spin-to-charge conversion in standard heterostructures, the standalone Pt film lacks an adjacent spin source. Crucially, the finite spin polarization induced within the Pt 5d band by the external magnetic field, originating from its exchange-enhanced paramagnetism, is orders of magnitude weaker than that of ferromagnets~\cite{Clogston64Interpretation, Soulen98Measuring}, and is thus insufficient to sustain the observed high-efficiency THz emission. Consequently, the ISHE is precluded as the primary generation mechanism in the single-layer Pt film.
	
	\begin{figure*}
		\includegraphics[width=0.8\linewidth]{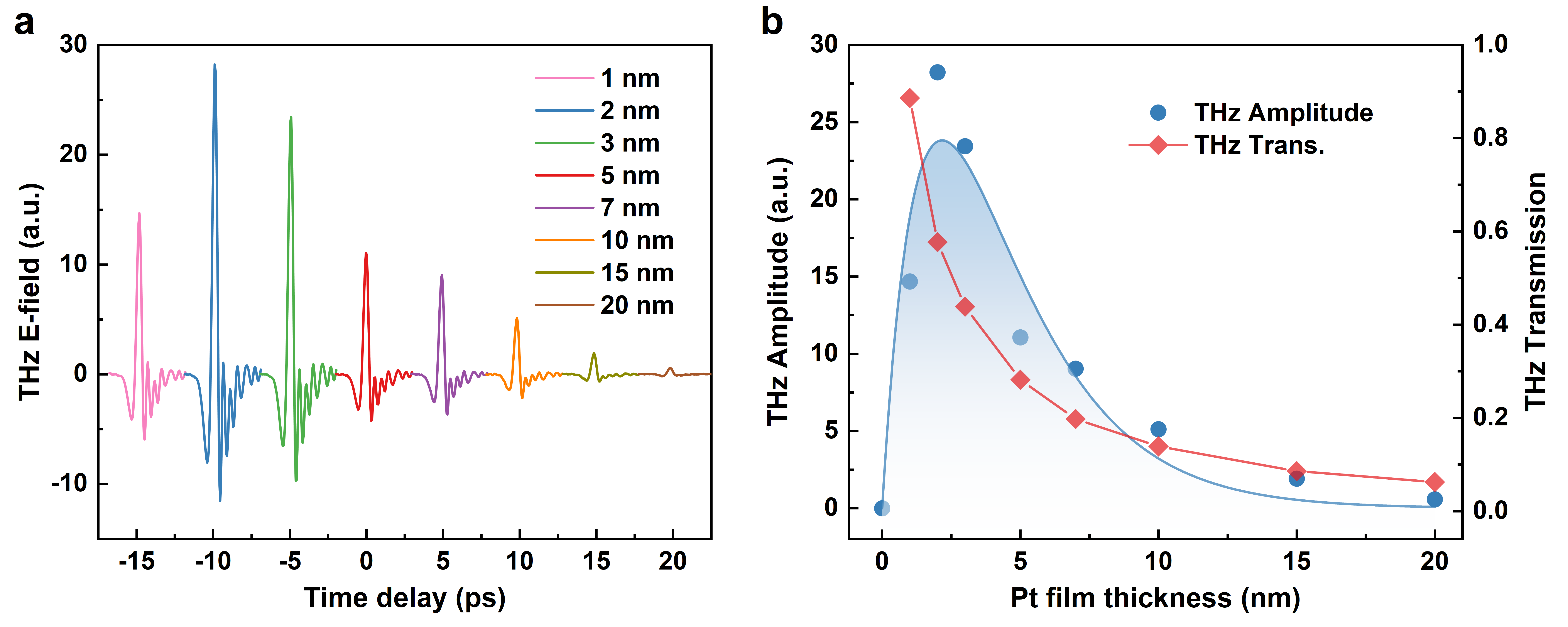}  \centering
		\subfigure{\label{4a}} 
		\subfigure{\label{4b}}
		\vspace{-0.3cm}
		\caption{\textbf{Thickness optimization of the Pt film.} \textbf{a} Time-domain THz waveforms emitted from Pt films of varying thicknesses, recorded at 10 K and 7 T. The transisents are horizontally offset for clarity. \textbf{b} Left axis: THz peak amplitude (circles) as a function of the Pt film thickness. The solid curve represents a fit accounting for the optical pump absorption and the effective THz conductive screening. Right axis: Independently measured THz transmittance (squares), exhibiting a monotonic decay with increasing film thickness.}	\label{4}
	\end{figure*}

	\vspace{0.1cm}	
	\noindent\textbf{Material dependence and Nernst-sign correlation.} Beyond the emission symmetries, the material dependence of the THz radiation provides a decisive signature of its physical origin. As presented in Fig. \ref{3a}, the THz amplitudes from a series of 5-nm-thick transition-metal films at 10 K and 7 T, exhibit an element-specific variation, accompanied by a polarity reversal of the THz electric field (Supplementary Fig. S5). Strikingly, the Pt film generates a strong positive transient, whereas W emits a pulse with comparable magnitude but inverted polarity. Ta replicates the reversed phase with an attenuated amplitude. Although the Pt-versus-W/Ta polarity inversion tracks the sign change of their spin Hall angles, the absence of interfacial spin injection excludes any ISHE-based interpretation. Instead, the distinctive material dependence indicates that an intrinsic bulk transport coefficient, possessing an identical sign correlation, drives the transient transverse charge current.
	
	The ultrafast PNE fulfills these rigorous symmetry constraints. Upon femtosecond pump excitation, the depth-dependent optical absorption establishes a transient out-of-plane electron temperature gradient (${\nabla T}_z$) across the ultrathin metallic film. Under an external magnetic field, the longitudinal thermal gradient induces a transverse charge current $\bm{j}_\mathrm{N} $ via the Nernst effect, $\bm{j}_\mathrm{N}\propto \sigma_{\mathrm{F}} Q_\mathrm{N} ({\nabla T}_z \times \bm{B}$), where $\sigma_{\mathrm{F}}$ and $Q_\mathrm{N}$ denote the electrical conductivity and the Nernst coefficient, respectively\cite{Feng23NL,LuWei2022ultrafast}. The transient current radiates a THz electric field ($\bm{E}_{\mathrm{THz}} \propto {\partial \bm{j}_\mathrm{N}}/{\partial t}$) that is polarized orthogonal to both ${\nabla T}_z$ and $\bm{B}$, fully accounting for the observed symmetries in Fig. \ref{2}. The opposite signs of the intrinsic low-temperature Nernst coefficients for Pt and W are independently verified via our steady-state magneto-transport measurements (Supplementary Fig. S6). The Nernst-sign alternation dictates the polarity reversal of the THz transients. This direct Nernst-phase correlation further confirms the PNE as the dominant generation mechanism in the standalone heavy-metal films.
		
	Quantitatively, the amplitude of the Nernst-mediated THz emission is determined by the product of the electrical conductivity $\sigma_{\mathrm{F}}$, the Nernst coefficient $Q_\mathrm{N}$, and the thermal gradient ${\nabla T}_z$. In low-conductivity metals such as Ta, Mn, and Ti, the small $\sigma_{\mathrm{F}}$ attenuates the THz output. For high-conductivity metals, $Q_\mathrm{N}$ determines the emission efficiency. Specifically, heavy metals like Pt and W possess strong spin-orbit coupling and complex $d$-band Fermi surfaces, which generates large Berry curvatures and sharp energy derivatives of the transverse conductivity near the Fermi level. These intrinsic topological features give rise to their enhanced Nernst response~\cite{Guo14Anomalous, Ahuja91PtFs, Behnia2016Nernst, Xiao10Berry}. In contrast, the trivial band topology inherent to simple $s$-band metals such as Cu enforces a vanishing Nernst coefficient \cite{Xiao10Berry}. In addition, the high thermal conductivity of Cu accelerates the spatial diffusion of hot carriers, thereby rapidly diminishing the driving thermal gradient and quenching the THz emission.
	
	As presented in Fig. \ref{3a}, a threefold enhancement in the THz peak amplitude is acquired from the sputtered Pt$_{0.8}$Ti$_{0.2}$ alloy relative to pure Pt. While heavy-metal alloying is widely employed to increase the spin Hall angle~\cite{Meinert20High,Wang23Enhancement,Zou25Enhanced,Janus25Enhanced}, the Ti-induced lattice disorder drastically suppresses the thermal conductivity \cite{Wang22Giant,Janus25Enhanced}. The restricted heat dissipation confines the deposited laser energy, directly steepening the transient out-of-plane temperature gradient. The magnified driving force thus overcompensates for the disorder-induced reduction in electrical conductivity, boosting the THz output. This underlying thermodynamics also account for the low saturation fluence extracted earlier (Fig. \ref{2d}). Within the two-temperature model, the linear dependence of the electronic heat capacity on the electron temperature limits the transient thermal response, compelling the temperature gradient to scale sublinearly at high pump fluences \cite{Eesley1986PRB, Lin2008PRB}. Owing to the suppressed thermal transport, the Pt-Ti alloy enters the saturation regime at a lower excitation density of 33.6 $\mu$J/cm$^2$  than pure Pt.
	
		\vspace{0.2cm}

	\noindent\textbf{Temperature evolution of the photo-Nernst emission.} As illustrated in Fig. \ref{3b}, the THz amplitudes for all three heavy-metal films (Pt, W, and Pt$_{0.8}$Ti$_{0.2}$) decrease abruptly from 10 K to 75 K, followed by a more gradual linear reduction up to 300 K, resulting in an order-of-magnitude attenuation (Supplementary Fig. S4c and S4d). The near-unity residual resistivity ratio of the sputtered films indicates a defect-dominated charge transport, enforcing a nearly temperature-independent electrical conductivity. Our two-temperature model simulations reveal that the peak out-of-plane electronic thermal gradient exhibits a variation of merely 6\% increase across the entire investigated temperature window (Supplementary Fig. S7). Moreover, the ubiquitous low-temperature  enhancement observed here contradicts the material-dependent Nernst response~\cite{Luo23Nernst, Chiang24Nernst}. Therefore, these steady-state transport coefficients fail to reproduce the massive signal enhancement at cryogenic temperatures. Far from equilibrium, electron-phonon scattering dominates the transient transport of photoexcited carriers \cite{Zhukov2006PRB, Groeneveld95}. As the lattice cools well below the Debye temperature, the severe depopulation of thermal phonons freezes out the scattering channel \cite{Allen87Theory, Groeneveld95}. The extended mean free path of the non-equilibrium carriers thus sustains an efficient transverse deflection, which yields the orders-of-magnitude higher effective mobility extracted earlier, resulting in the low-temperature THz enhancement.
	
	The temperature evolution of the PNE is distinct from that of the ISHE in FM/NM bilayers and the ANE in single ferromagnetic films (Supplementary Fig. S8). In standard FM/Pt heterostructures, the THz emission is attenuated upon cooling, which originates from the resistivity-dependent scaling of the spin Hall angle and the reduction in interface spin transparency \cite{Matthiesen20Temperature,Sandeep22Large,Kumar23THz}. Likewise, despite sharing a transverse thermoelectric origin, the ANE in the metallic ferromagnets displays only a marginal low-temperature THz enhancement, constrained by the vanishing intrinsic Nernst coefficient as the temperature approaches absolute zero \cite{GAUTAM2018264,De21Temperature,Nagaosa10AHE}. In contrast, the order-of-magnitude cryogenic enhancement in standalone heavy-metal films constitutes a defining hallmark of the ultrafast PNE, wherein the extended momentum relaxation time and the resulting surge in hot-carrier mobility amplify the non-equilibrium Nernst response. The contrasting thermal signatures establish the ultrafast PNE as a bulk THz emission mechanism, independent of interfacial spin injection and magnetic ordering.
	
	\vspace{0.2cm}
	\noindent\textbf{Thickness optimization of the Pt film.}  Having established the ultrafast PNE mechanism, the THz emission is further optimized by measuring the thickness dependence of single-layer Pt films from 1 to 20 nm at 10 K and 7 T. As shown in Fig. \ref{4a}, the THz amplitude shows a pronounced maximum at a thickness of 2 nm and subsequently undergoes a rapid attenuation, becoming negligible beyond 20 nm. While conventional spintronic heterostructures display a similar thickness dependence, their optimal thickness is dominated by spin transport parameters, namely the spin diffusion length and interfacial spin-mixing conductance~\cite{Seifert2016Efficient, Cheng21Studying, Seifert22Spintronic, Kumar23Ultrafast}. Conversely, the thickness scaling of the standalone Pt film is emerges from the competition between the optical pump absorption driving the transient electron temperature gradient (Supplementary Fig. S7) and the conductive screening of the generated THz field. The trade-off is quantitatively described by the relation $E_{\mathrm{THz}} \propto (1-e^{-\alpha d}) e^{-\beta d}$, where $\alpha$ is the optical absorption coefficient, $d$ is the film thickness, and $\beta$ represents the THz attenuation coefficient. An effective THz screening length (1/$\beta$) of 2.6 nm is extracted from the fit to the experimental data (solid curve in Fig. \ref{4b}). The severe attenuation in thicker films is corroborated by THz transmittance measurements (Fig. \ref{4b} and Supplementary Fig. S9), which undergoes a rapid monotonic decay, falling below 5\% as the film thickness approaches 20 nm. Accordingly, scaling the 5-nm Pt$_{0.8}$Ti$_{0.2}$ alloy down to the optimal thickness suppose to further enhance the single-layer emission, reaching an amplitude equivalent to that of benchmark spintronic bilayers.
		
	In conclusion, efficient THz radiation from standalone heavy-metal nanofilms is established under low-temperature and high-magnetic-field conditions. Through a comprehensive analysis of the emission symmetries, material dependence, and thermal evolution of the THz transients, the ultrafast photo-Nernst effect is determined as the governing generation mechanism. Moreover, the single-layer emission efficiency is substantially enhanced through thickness scaling and alloying-induced thermal engineering, reaching an amplitude comparable to that of standard spintronic bilayers. This work establishes standalone heavy metals as active THz emitters, providing a contact-free, all-optical methodology for probing non-equilibrium magneto-thermoelectric dynamics across diverse spintronic and quantum materials.

		\setlength{\parindent}{0pt}
	\section{Experimental method}
	\textbf{Sample fabrication:} All metallic thin films were deposited at room temperature by magnetron sputtering in a 3 mTorr Ar atmosphere using an ultrahigh-vacuum AJA sputtering system with a base pressure below $3 \times 10^{-9}$ Torr. Double-side-polished MgO substrates with a thickness of 0.5 mm were utilized, and all films were capped with an \textit{in-situ} grown MgO(2 nm)/W(1 nm) bilayer to prevent oxidation. The thicknesses of the metallic layers were controlled by the deposition time, based on sputtering rates precalibrated via X-ray reflectometry. The structure characterization and electrical transport measurements are detailed in Section S2 of the Supporting Information.
		
	\medskip
	\textbf{THz emission setup:} A Ti:sapphire regenerative amplifier (central wavelength of 800 nm, pulse duration of 35 fs, repetition rate of 1 kHz) was employed as the optical source. The collimated pump beam with a beam diameter of 11 mm was directed onto the sample at normal incidence. The emitted THz transients were collected and focused by an off-axis parabolic mirror onto a 2-mm-thick ZnTe crystal for electro-optic sampling. A high-resistivity silicon wafer was inserted to block the residual pump pulses. A wire-grid THz polarizer was utilized to select the specific THz polarization component (horizontal or vertical). The 10$\times$10 mm$^2$ sample was mounted on an 8-mm-diameter holder and positioned inside a split-coil superconducting magnet cryostat, enabling independent control over the sample temperature and the external magnetic field. All THz emission measurements were performed under ambient atmospheric conditions. Further details regarding the optical beam path are provided in Fig. S1 of the Supporting Information.
	
	\medskip
	\textbf{Supplementary information} \par Supplementary materials are available for this article online.
	
	\medskip
	\par \textbf{Data Availability} \par The data that support the findings of this study are available from the corresponding author upon reasonable request.
	
	\medskip
	\textbf{Acknowledgements} \par 
	This work was supported by the National Key Research and Development Program of China (Grants No. 2024YFA1611300 and No. 2022YFA1402600), the Beijing Natural Science Foundation (Grants No. Z230006 and No. IS25044) and the National Natural Science Foundation of China (Grants No. 12574349, No. 12274438 and No. 12504156). E.E.M.C. acknowledges support from the Singapore Ministry of Education (MOE) Academic Research Fund Tier 3 (MOE-MOET32023-0003) grant. This work was supported by the Synergetic Extreme Condition User Facility (SECUF, https://cstr.cn/31123.02.SECUF).
	
	\medskip
	\bibliographystyle{apsrev4-2}
	\bibliography{Pt_THz_0304}
	
\end{document}